\documentclass[preprint2]{aastex}

 \newcommand{\cc}{\c{c}}
 
 \newcommand{\coes}{\cc\~{o}es}

\begin{document}

\title{A study of the evolution of the accretion disk \\ of V2051 Oph through 
       two outburst cycles\thanks{Based on observations made at the 
       Laborat\'orio Nacional de Astrof\'{\i}sica, CNPq, Brazil.}}

\author{R. Baptista\altaffilmark{1,2}}
\email{bap@astro.ufsc.br}
\author{R. F. Santos\altaffilmark{1}}
\and
\author{M. Fa\'undez-Abans\altaffilmark{3} and A. Bortoletto\altaffilmark{3,4}}

\altaffiltext{1}{Departamento de F\'{i}sica , Universidade Federal de Santa 
    Catarina, Campus Trindade, 88040-900, Florian\'opolis, SC, Brazil}
\altaffiltext{2}{SOAR Telescope, Colina El Pino s/n, Casilla 603, La Serena, 
    Chile}
\altaffiltext{3}{Laborat\'{o}rio Nacional de Astrof\'{i}sica, Rua Estados 
    Unidos 154, 37504-364, Bairro das Na{\coes}, Itajub\'{a}, MG, Brazil}
\altaffiltext{4}{Instituto de Astronomia, Geof\'isica e Ci\^encias
     Atmosf\'ericas, Universidade de S\~ao Paulo, Rua do Mat\~ao 1228,
     05508-900, S\~ao Paulo, SP, Brazil}

\begin{abstract}
 We follow the changes in the structure of the accretion disk of the dwarf 
 nova V2051~Oph along two separate outbursts in order to investigate the 
 causes of its recurrent outbursts.
 We apply eclipse mapping techniques to a set of light curves covering
 a normal (July 2000) and a low-amplitude (August 2002) outburst to derive 
 maps of the disk surface brightness distribution at different phases along
 the outburst cycles.
 The sequence of eclipse maps of the 2000 July outburst reveal that the 
 disk shrinks at outburst onset while an uneclipsed component of 13 per 
 cent of the total light develops. The derived radial intensity 
 distributions suggest the presence of an outward-moving heating wave 
 during rise and of an inward-moving cooling wave during decline.
 The inferred speed of the outward-moving heating wave is $\simeq 1.6\;
 kms^{-1}$, while the speed of the cooling wave is a fraction of that.
 A comparison of the measured cooling wave velocity on consecutive nights 
 indicates that the cooling wave accelerates as it travels towards disk
 center, in contradiction with the prediction of the disk instability 
 model. From the inferred speed of the heating wave we derive a viscosity
 parameter $\alpha_{hot} \simeq 0.13$, comparable to the measured 
 viscosity parameter in quiescence.  
 The 2002 August outburst had lower amplitude ($\Delta B \simeq 0.8$ mag) 
 and the disk at outburst maximum was smaller than on 2000 July.  
 For an assumed distance of 92 pc, we find that along both outbursts the 
 disk brightness temperatures remain below the minimum expected according 
 to the disk instability model.
 The results suggest that the outbursts of V2051 Oph are caused by bursts
  of increased mass transfer from the mass-donor star.
\end{abstract}


\section {Introduction} \label{intro}

Cataclysmic variables (CVs) are close interacting binaries in which a 
late-type star (the secondary) overfills its Roche lobe and loses matter to 
a white dwarf companion via an accretion disk or co\-lumn (Warner 1995).  
The subclass of dwarf novae comprises low-mass transfer CVs in which mass 
is fed to a weakly magnetic ($\left| \vec{B} \right| \leq 10^{6} G$) white 
dwarf. These binaries show recurrent outbursts of up to 5 magnitudes on 
timescales of weeks to months as a consequence of a sudden increase in mass 
inflow in the accretion disk. 

There are two competing models to explain the cause of the sudden increase 
in mass accretion. In the disk-instability model (DIM), matter is 
transferred at essentially constant rate to a low viscosity disk and 
accumulates in an annulus until a thermal-viscous instability switches the
disk to a high viscosity regime and the gas diffuses rapidly inward and onto
the white dwarf. In this case, the outburst starts at the radial position
where the thermal instability first occurs and propagates as a heating wave
on the rise and as a cooling wave along the decline back to quiescence 
(Hameury et~al. 1998, Lasota 2001). 
In the mass transfer instability model (MTIM), the outburst is the 
time-dependent response of a high viscosity accretion disk to a burst of 
enhanced mass transferred from the secondary star (Bath 1975). This model 
predicts that the disk shrinks at the onset of the outburst in response to 
the sudden addition of matter with low angular momentum, and 
that the disk viscosity in quiescence and in outburst are similar.
DIM predicts no reduction of disk radius at outburst start but demands that
the viscosity parameter in quiescence be 5-10 times smaller than the 
viscosity in outburst. 
The thermal limit cycle of DIM also predicts that the outbursting parts of 
the accretion disk should be hotter than a critical temperature 
$T_\mathrm{eff}$(crit) (Warner 1995).
Furthermore, the inward cooling wave is expected to 
decelerate as it travels towards disk center (Menou et~al. 1999).
Thus, there are several predictions that can be tested from observations of 
accretion disks through outburst cycles with the aid of indirect imaging
techniques such as eclipse mapping (Horne 1985). 

The interest in testing both models against observations has somehow 
reduced over the last two decades as a consequence of a wide acceptance of 
DIM as the correct explanation. 
Two arguments have been key in setting the dominance of DIM.
They are based on the assumption that the matter transferred from the 
secondary star is deposited at the disk rim, where the gas stream hits the 
accretion disk to form a bright spot (BS). Given this assumption, one may 
predict that (i) a burst of enhanced mass transfer rate would inevitably 
lead to an increase in the luminosity of the BS and that (ii) an MTIM-driven
outburst can only lead to outside-in outbursts because the additional matter
is always deposited at the disk rim. The existence of inside-out
outbursts and the lack of observational support for an increase in BS 
luminosity at outburst onset seem to argue against MTIM. 
However, one should note that both arguments (and the reasoning to drop
MTIM) fall apart if the assumption upon which they rely is incorrect.
As we shall see later (Sect.\,5), this may indeed be the case.

V2051 Oph is an ultra-short-period CV ($P_{orb} < 2\;hr$), almost all of 
which are either polars with highly magnetic white dwarfs ($\left| \vec{B} 
\right| \sim 10^{7} G$) or members of the SU UMa subclass of 
superoutbursting dwarf novae (Warner 1995). The binary was discovered by 
Sanduleak (1972) and since then several classifications were proposed. 
Warner \& O'Donoghue (1987) listed some characteristics of the object that 
distinguish it among other CVs: with its $P_{orb} = 90$ min it has one of 
the shortest periods known; it is one of the few short-period CVs to show 
deep eclipses ($B\simeq 2.5\,mag$); unlike other eclipsing systems, it 
possesses strong flickering (random brightness fluctuations of 0.1-1 
magnitudes on timescales from seconds to minutes) and flare activity which 
produce a wide assortment of eclipse morphologies (Warner \& Cropper 1983). 
V2051~Oph was proposed to be a low-field polar system by Warner \& 
O'Donoghue (1987) based on the interpretation of their eclipse maps and on 
the observation of a 42-s oscillations in the optical during outburst, 
reminiscent of the rapid oscillation seen in polars. A consensus was reached 
with the observation of a superoutburst during which superhumps were 
detected (Kiyota \& Kato 1998, Vrielmann \& Offutt 2003), establishing its 
classification as an SU~UMa-type dwarf novae. 

As part of a long-time campaign to study the flickering properties of a 
sample of eclipsing CVs (e.g., Baptista \& Bortoletto 2004), we happened to 
observe V2051~Oph in outburst on two occasions.  This paper reports the 
analysis of the light curves of V2051~Oph collected along these two 
outbursts with eclipse mapping techniques.  The resulting maps deliver the 
disk brightness distribution in quiescence, on the rise to maximum, during
maximum light and in the decline phase.  Section 2 describes the observations
and Sect.\,3 gives the details of the data analysis. The results are 
presented in Sect.\,4 and discussed in Sect.\,5 in a comparison with
predictions from the dwarf nova outburst models.

\section{The observations}

Time-series of high-speed CCD photometry of V2051 Oph were obtained with a 
CCD camera ($385\times 578$ pixels, 0.58 arcsec/pixel) attached to the 1.6m 
telescope of Laborat\'orio Nacional de Astrof\'{i}sica (LNA), in Itajub\'a,
Brazil, between 2000 July 28 and August 2 in B and V bands, and on 2002 
August 4, 5 and 7 in the B band. The observations are summarized in 
Table~\ref{observa}. The second column lists the
starting time of the observations in Julian Date and the third column 
shows the time resolution in seconds ($\Delta t$). The fifth column lists 
the eclipse cycle number (E); the numbers in parenthesis indicate 
observations that, because of interruptions caused by, p.ex., clouds, do 
not cover the eclipse itself.  The sixth column indicates the 
binary phases covered by the observations (running from $-0.5$ to +0.5 
with eclipse center at phase 0.0), the seventh column lists the number of 
points in the light curve ($N_p$).  The eighth column gives an estimate of 
the quality of the observations.  The seeing ranged from 1.0 to 2.2 arcsec.
Part of the data (quiescence) was presented in Baptista \& Bortoletto (2004), 
but is repeated here for completeness. 
All light curves were obtained with the same instrumental set and 
telescope, which ensures a high degree of uniformity to the data set.
%

Data reduction procedures included bias subtraction, flat-field correction,
cosmic rays removal and aperture photometry extraction. Time series were
constructed by computing the magnitude difference between the variable and a
reference comparison star with scripts based on the aperture photometry
routines of the APPHOT/IRAF package\footnote{IRAF is distributed by National
  Optical Astronomy Observatories, which is operated by the Association of
  Universities for Research in Astronomy, Inc., under contract with the
  National Science Foundation.}.  Light curves of other comparison stars in 
the field were also computed in order to check the quality of the night and 
the internal consistency and stability of the photometry over the time span 
of the observations.  

We estimate that the absolute photometric accuracy of these observations is
about 10 per cent. On the other hand, the analysis of the relative flux of 
the comparison star of all observations indicates that the internal error 
of the photometry is less than 2 per cent. The individual light curves have 
typical signal-to-noise ratios of $S/N=40-50$ out-of-eclipse and $S/N=10-20$ 
at mid-eclipse.  Additional details of the data reduction procedures are 
given in Baptista \& Bortoletto (2004).  

\begin{figure}
\epsscale{1.05}
\plotone{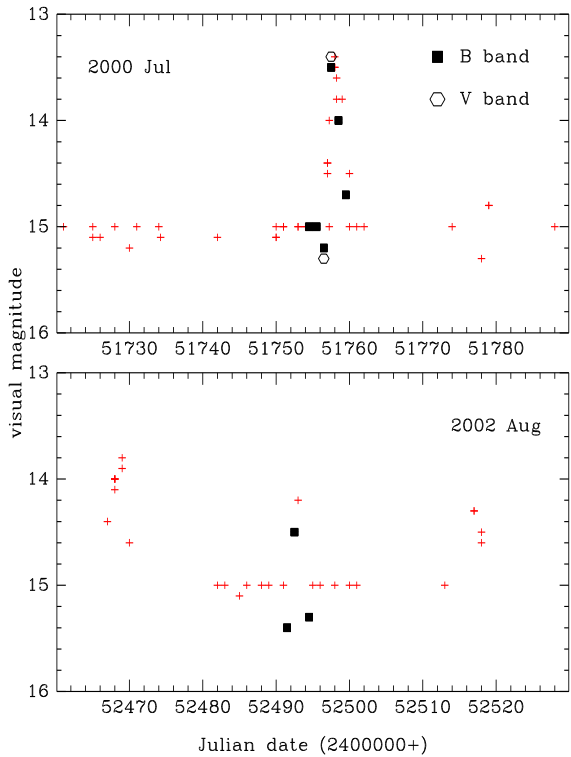}
\caption[c]{Historical light curve of V2051 Oph around the epoch of the two
  outbursts, constructed from observations made by the AAVSO (crosses). 
  Average out-of-eclipse B and V magnitudes from our data are indicated 
  respectively by filled squares and open symbols. The B band magnitudes
  were displaced vertically by $-0.5$ mag for visualization purposes. }
\label{fig1}
\end{figure}

The historical light curve of V2051~Oph 
\footnote{based on data from the AAVSO international database} 
around the epoch of both outbursts is shown in Fig.\,\ref{fig1}. The star 
went in outburst between 2000 July 30 and 31 (about JD 2451758) 
[Fig.\,\ref{fig1}, top panel]. This was a normal outburst, with an 
amplitude of 2 magnitudes and a time span of 2 days.  Our observations 
cover three days before maximum, the outburst maximum, and two days along 
the decline phase.  The star was also seen in outburst on 2002 July 9-10 
(Fig.\,\ref{fig1}, bottom panel).  We did not observe this outburst.  Our 
observations at this epoch cover the following, low-amplitude ($\simeq 0.8$ 
mag) and short outburst on 2002 August 5 (JD 2452493). The runs on 2002 
Aug 4 and 7 frame, respectively, the day before outburst and a late decline
stage when the star was almost back to its previous quiescent brightness
level.  A similar, low-amplitude outburst occurred 20 days after the one 
recorded in our data.
%

Average out-of-eclipse B and V magnitudes measured from our light curves
are superimposed on the AAVSO historical data (Fig.\,\ref{fig1}). The B-band 
magnitudes were displaced vertically by $-0.5$ mag for visualization 
purposes. Only B-band data is available for the 2002 outburst.  The V-band 
magnitudes are consistent with the visual observations. However, the B-band 
magnitudes are systematically fainter than both the V-band and visual 
contemporary measurements, implying a color index $(B-V) \simeq +0.5$ mag. 
This is noticeably redder than the color of V2051~Oph previously reported 
in the literature, $-0.1 \leq (B-V) \leq +0.2$ (Bruch 1983; Vogt 1983), and
may be associated to the extended state of lower brightness the object has 
gone through in 1996 (see Baptista et~al. 1998a). The color index changed
from $(B-V)=+0.4$~mag to $(B-V)=+0.6$~mag from quiescence to outburst 
maximum in 2000.

\section{Data analysis}

\subsection{Light curve construction} \label{lightcurve}

\begin{figure*}
\includegraphics[bb=2cm 1cm 18cm 24cm, angle=270, scale=0.7]{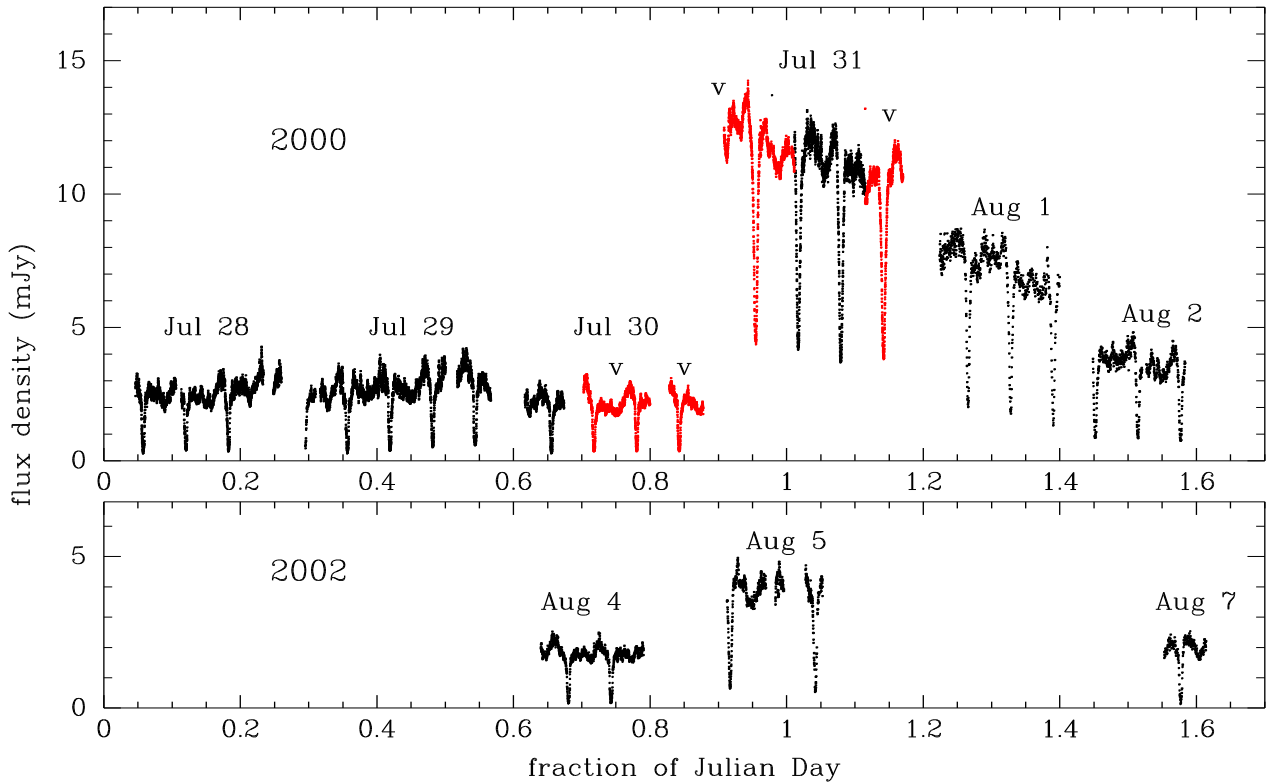}
\caption[c]{Light curves of V2051 Oph through the 2000 July (top) and 2002 
  August (bottom) outbursts.  The separation of data from consecutive nights 
  is arbitrarily reduced to 0.3~d for visualization purposes.  Labels above 
  the data of each night indicate the date. The V-band light curves are shown
  with gray symbols and are indicated by a small 'v' label above the data.  
  The V-band fluxes are arbitrarily scaled by a factor 0.75 also for 
  visualization purposes. }
\label{fig2}
\end{figure*}

The B- and V-band individual light curves of V2051~Oph are shown in 
Fig.~\ref{fig2}, with arbitrary adjustments in the x-axis separation and 
V-band fluxes for visualization purposes.
%

\begin{figure*}
\includegraphics[angle=270, scale=0.6]{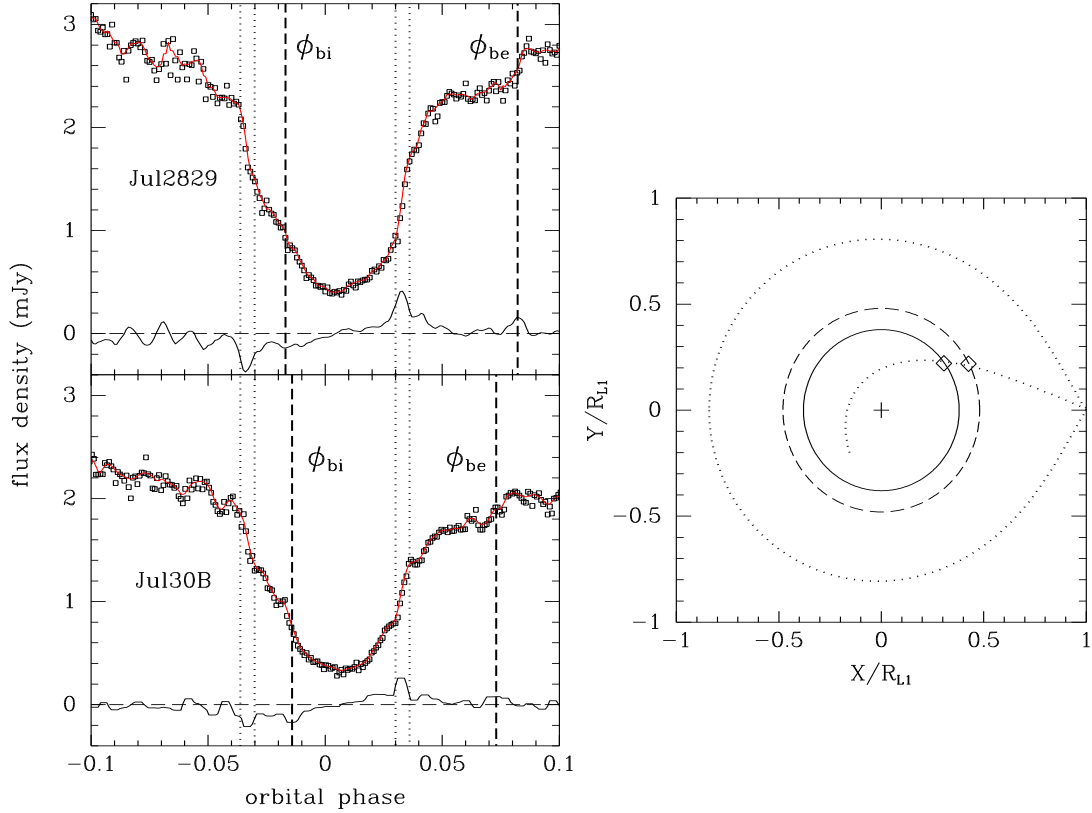}
\caption[c]{Left: The Jul2829 (top) and Jul30B (bottom) light curves and 
  corresponding derivative curves at a phase resolution of 0.001~cycle.
  The median filtered version of the light curve is shown as a solid grey 
  line in each case. Vertical dotted lines indicate the contact phases of 
  the WD and vertical dashed lines mark the mid-ingress ($\phi_{bi}$) and 
  mid-egress ($\phi_{be}$) phases of the BS. Right: the binary geometry for 
  $q=0.19$.  Dotted lines depict the primary Roche lobe and the ballistic 
  stream trajectory.  The WD is shown as a cross; the BS positions are 
  marked by diamonds and the corresponding disk sizes are indicated by 
  circles of radii $0.48\;R_{L1}$ (dashed line) and $0.38\;R_{L1}$ (solid 
  line), respectively for Jul2829 and Jul30B. }
\label{fig3}
\end{figure*}

The data were grouped per passband and outburst stage and combined to 
produce light curves of improved S/N and reduced flickering influence.
The B-band data from 2000 July 28 and 29 were combined into a single 
average light curve representative of the quiescence state (hereafter 
Jul2829) because they are at the same brightness level and there is no 
perceptible difference in light curve morphology from one night to the 
other.  The B-band data from 2000 July 30 (Jul30B) was treated separately 
as it shows a decrease in brightness with respect to the data from the 
previous two nights as well as changes in eclipse shape (Fig.~\ref{fig3}).  
There is a continuous decrease in brightness along the observations on 
2000 Jul 31 and Aug 1.  The individual light curves of those nights were 
scaled (by factors $\la 10$ per cent) to a mean, common out-of-eclipse flux 
level before combining them. For both nights it was possible to obtain a 
good match of the eclipse shape and out-of-eclipse level for all combining 
curves.

For each light curve, we divided the data into phase bins of 0.003 cycle 
and computed the median flux at each bin to reduce the influence of 
flickering.  The median of the absolute deviations with respect to the 
median flux was taken as the corresponding uncertainty for each bin. The 
light curves were phase-folded according to the linear ephemeris (Baptista 
et~al. 2003),
\begin{equation}
T_{mid}=BJDD\; 2\,443\,245.97752 + 0.062\,427\,8634 \times E, 
\label{efem}
\end{equation}  
where BJDD denotes Barycentric Dynamical Time. A small phase corrections 
of $+0.0061$~cycle was further applied to the 2002 data to remove the 
long-term cyclical period changes (Baptista et~al. 2003) and to make the 
center of the white dwarf eclipse coincident with phase zero.

The white dwarf (WD) and bright spot (BS) eclipses are seen as sharp 
changes in the slope of the eclipse shape in quiescence light curves (e.g., 
Fig.~\ref{fig3}).  We separated the contribution of the WD from the Jul2829, 
 Jul30B and Aug04 average light curves with a light-curve decomposition
technique (Wood et~al. 1985) in order to compute eclipse maps of only the 
accretion disk.  A zoom of the Jul2829 and Jul30B light curves
around eclipse is shown in Fig.~\ref{fig3}. 
%
%
The original light curve is smoothed with a median filter of width 0.006 
cycle (the phase width of the WD ingress/egress feature) and its numerical 
derivative is calculated.  The derivative curve is smoothed with the same 
median filter as above to reduce noise and improve the detection of the WD
and BS features.  The WD and BS ingress/egress features are seen as those 
intervals where the derivative is significantly different from zero.  A 
spline function is fitted to the remaining regions in the derivative to 
remove the contribution of the slowly varying eclipse of the extended disk, 
and estimates of the WD flux are obtained by integrating the 
spline-subtracted derivative curve at ingress and egress.  The light curve 
of the WD is reconstructed by assuming that its flux is zero between
ingress and egress and constant outside eclipse. The WD contribution is 
then removed from the data by subtracting the reconstructed WD curve from 
the original light curve.  The eclipse maps of the Jul2829, Jul30B and 
Aug04 data shown in Sect.~\ref{mapas} were computed using WD-subtracted 
light curves. 

A close inspection on the light curves in Fig.~\ref{fig3} reveal 
that, while the phase width of the WD ingress/egress feature is the same 
on both nights, the integrated WD flux is reduced by $\simeq 25$ per cent 
on Jul30B.  This could be caused by an increased optical depth of the inner
disk gas or by a thicker disk rim on that night, resulting in a lower
visible fraction of the WD surface.

The average light curves are shown in the left panels of Figs.~\ref{fig4}, 
\ref{fig5} and \ref{fig6}. The Jul2829 and the Aug05 light curves have 
flares at $\phi \sim -0.11$~cycle.  The error bars around these flares were 
artificially increased to minimize their influence on the eclipse mapping 
modeling.

\subsection{Eclipse mapping} \label{mem}

The eclipse-mapping method (Horne 1985) is an inversion technique that 
uses the information in the eclipse shape to reconstruct the disk surface 
brightness distribution.  The reader is referred to Baptista (2001) for a 
review on the subject. Eclipse mapping techniques (Baptista \& Steiner 
1993) were applied to the average light curves of Sect.~\ref{lightcurve} to 
solve for a map of the disk surface brightness plus the flux of an additional
uneclipsed component in each case.  The uneclipsed component accounts for all
light that is not contained in the eclipse map (i.e., light from the 
secondary star, a vertically extended disk wind, or both). The reader is 
referred to Rutten et~al. (1992a) and Baptista et~al. (1996) for the details
of and tests with the uneclipsed component.

All variations in a light curve in the standard eclipse mapping method are
interpreted as being caused by the changing occultation of the emitting 
region by the secondary star.  Thus, any out-of-eclipse brightness change
(e.g., the orbital hump caused by anisotropic emission from the BS) has to
be removed before applying the technique to a light curve. This was done by
fitting a spline function to the phases outside eclipse, dividing the light
curve by the fitted spline, and scaling the result to the spline function
value at phase zero in each case. This procedures removes orbital
modulations with minimal effects on the eclipse shape itself.

Our eclipse map is a flat Cartesian grid of $75\times75$ pixels centered on 
the primary star with side $2 R_{L1}$ (where $R_{L1}$ is the distance from 
the disk center to the inner Lagrangian point L1). The eclipse geometry is 
defined by the mass ratio $q$ and the inclination $i$, and the scale of the 
map is set by $R_{L1}$.  We adopted the values of Baptista et~al. (1998a), 
$R_{L1}=0.422\; R_\odot$, $q=0.19$ and $i=83\degr$, which correspond to a 
white dwarf eclipse width of $\Delta\phi = 0.0662$~cycle. This combination
of parameters ensures that the white dwarf is at the center of the map.
The reconstructions were performed with a polar gaussian default function
(Rutten et~al. 1992a) with radial blur width $\Delta r = 0.0266\;R_{L1}$ and
azimuthal blur width $\Delta\theta = 30\degr$.  
The reconstructions reached a final $\chi^{2}$ near or equal 1 
for all light curves.     

The statistical uncertainties in the eclipse maps are estimated with a 
Monte Carlo procedure (see, e.g., Rutten et~al. 1992a). For a given input 
data curve, a set of 20 artificial light curves are generated in which the
data points are independently and randomly varied according to a Gaussian
distribution with standard deviation equal to the uncertainty at that point.
The artificial curves are fitted with the eclipse-mapping algorithm to 
produce a set of randomized eclipse maps.  These are combined to produce
an average map and a map of the residuals with respect to the average, 
which yields the statistical uncertainty at each pixel.  A map of the
statistical significance (or the inverse of the relative error) is obtained
by dividing the true eclipse map by the map of the standard deviations. 
The uncertainties obtained with these procedures are also used to estimate
the errors in the derived radial intensity and temperature distributions.

\section{Results}

\subsection{Bright spot phases and disk radius changes} \label{radius}

We used the derivative technique of Sect.~\ref{lightcurve} to measure 
the mid-ingress/egress phases of the BS ($\phi_{bi},\phi_{be}$) and to 
estimate the disk radius under the assumption that the BS is located where 
the stream of transferred matter hits the edge of the accretion disk.  
$\phi_{bi}$ and $\phi_{be}$ are taken as the phases of, respectively, 
minimum and maximum of the features corresponding to BS ingress/egress in 
the smoothed and spline-subtracted derivative of the light curve.

BS mid-ingress/egress phases for the light curves Jul2829 and Jul30B are 
shown as vertical dashed lines in the left-hand panels of Fig.~\ref{fig3}.  
We find $\phi_{bi}= -0.0170 \pm 0.0015$ and $\phi_{be}= +0.0822 \pm 0.0006$ 
for Jul2829, and $\phi_{bi}= -0.0142 \pm 0.0007$ and $\phi_{be}= +0.073 \pm 
0.001$ for Jul30B.  
BS enters eclipse {\em later} and reappears from eclipse {\em earlier} on 
Jul30B, indicating that the disk radius was smaller on that night.

Each pair of ($\phi_{bi},\phi_{be}$) values maps into a single position in 
the orbital plane for an assumed binary geometry ($i,q$). The x-y positions 
corresponding to the measured ($\phi_{bi},\phi_{be}$) values are indicated 
by diamonds in the binary diagram shown in the right-hand panel of 
Fig.~\ref{fig3}. The measured BS positions consistently fall along the gas 
stream trajectory for the assumed mass ratio $q=0.19$.
The circles that pass through these points are $R_{bs}= (0.480 \pm 0.015)\; 
R_{L1}$ and $R_{bs}= (0.380 \pm 0.015)\; R_{L1}$, respectively for Jul2829 
and Jul30B. The difference between the two values is formally significant 
at the 6-$\sigma$ confidence level. 
Thus, there is evidence that the disk shrank at outburst onset (i.e., the
night before outburst maximum), in agreement with the expectations of MTIM. 

It shall be noticed that this result comes mainly from one light curve 
(Jul30B). Given the large flickering amplitude of V2051~Oph, one cannot
exclude the possibility that the BS eclipse phases in this single light 
curve are affected by flickering.  We attempted to strengthen the Jul30B
result by repeating the analysis for the V-band light curves of the same
night. While the measured $\phi_{bi}$ is in good agreement with the B-band 
measurement, $\phi_{be}$ is unfortunately lost in the flickering on a slow 
egress slope.  A comparison of the eclipse shape of the B- and V-band light
curves of 2000 Jul 30 indicates that the BS is less compact in the V-band
than in the B-band.

We further tested whether the reduction in brightness and radius of Jul30B
was a common, short-lived effect caused by rapid changes in mass transfer
rate (on timescales shorter than that of an outburst). We measured BS eclipse
phases in the individual light curves of the previous nights, while the 
star was in quiescence.  We took particular attention to the first two
eclipses on 2002 Jul 28, for which the star had a brightness level comparable
to that of the Jul30B light curve. In all cases the BS eclipse phases (and
inferred disk radius) are consistent with the measurements obtained when we
combine all data taken on 2000 Jul 28-29. While the out-of-eclipse flux 
varies by $\simeq 20$ per cent among the individual light curves, we find
no evidence of decrease in disk radius nor correlation of disk radius with
out-of-eclipse brightness.  This suggests that the observed reduction in
disk radius reflects a change in disk structure particular of that night.

\subsection{Accretion disk structure}
\label{mapas}

\subsubsection{The 2000 Jul Outburst}

\begin{figure}
\includegraphics[bb=1.8cm 0.5cm 15.5cm 25cm, scale=0.57]{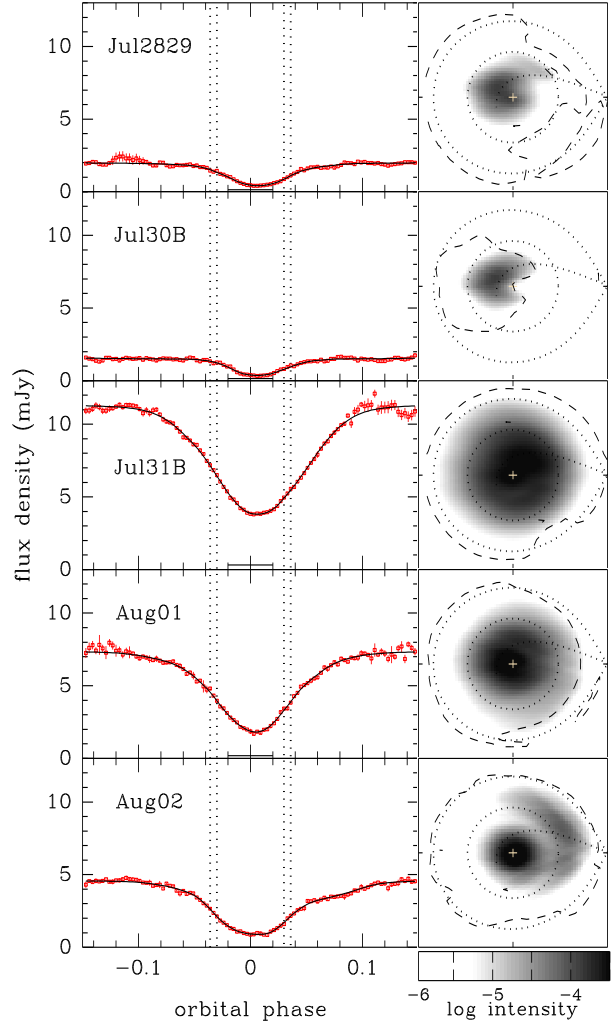}
\caption[c]{B-band light curves and eclipse maps along the 2000 Jul outburst:
  in quiescence (Jul2829), at outburst onset (Jul30B), outburst maximum 
  (Jul31B) and during decline (Aug01 and Aug02). Left: median light curves 
  (dots with error bars) and corresponding eclipse mapping models (solid 
  lines). Vertical dotted lines mark the contact phases of the WD eclipse. 
  Horizontal ticks depict the uneclipsed flux in each case.  Right: the 
  corresponding eclipse maps in a logarithmic greyscale common to all maps. 
  Brighter regions are indicated in black, fainter regions in white. A cross 
  marks the center of the disk; dotted lines show the primary Roche lobe, 
  the gas stream trajectory for $q=0.19$ and a reference disk of radius 
  $R_{bs}= 0.48\; R_{L1}$; the secondary is to the right of each map. The 
  horizontal bar indicates the logarithmic intensity level of the grey 
  scale. A dashed contour line is overploted on each eclipse map to 
  indicate the 3-$\sigma$ confidence level region. }
\label{fig4}
\end{figure}

B-band median light curves (dots with error bars) and eclipse mapping 
model curves (solid lines) are presented in the left panels of 
Fig.~\ref{fig4}.  The right panels shows the corresponding eclipse maps 
in a common logarithmic greyscale. 
The sequence of eclipse maps allows us to trace changes in the accretion 
disk structure at five different occasions along the outburst: in 
quiescence (Jul2829), and at outburst onset (Jul30B), maximum (Jul31B) 
and on two consecutive nights along the decline (Aug01 and Aug02). 
%

The light curve and eclipse map in quiescence are very similar to those of 
Baptista \& Bortoletto (2004), with enhanced emission along the gas stream 
trajectory inwards of the BS position.  The tip of the asymmetric emission
is consistent with the position where the ballistic stream trajectory
intercepts the $0.48\;R_{L1}$ quiescent disk radius.  The disk becomes 
fainter, smaller and largely asymmetric on Jul30B, with emission only 
along the gas stream trajectory. The lack of emission from the inner disk
regions suggest a reduction of accretion onto the WD, possibly as a
consequence of redistribution of matter and angular momentum in the disk
following outburst onset.  

The disk brightness distribution changes significantly on the timescale of 
one day, from outburst onset (Jul30B) to outburst maximum (Jul31B). The 
wide, V-shaped and fairly symmetric eclipse of Jul31B maps into a broad
brightness distribution that fills the primary Roche lobe, with a small
asymmetry towards the trailing side of the disk (the one containing the 
stream trajectory). The brightness distribution is not concentrated in
the inner disk regions as it would be expected for an opaque steady-state 
disk following the $T(R)\propto R^{-3/4}$ law (e.g., Frank et~al. 2002).
The eclipse becomes progressively narrower and U-shaped along the following
nights (Aug01 and Aug02).  The corresponding brightness distributions 
show the inward cooling and fading of the outer disk regions whereas the
inner disk remains at the same brightness level of outburst maximum.  
On Aug02 the outer disk is faint enough that an asymmetric and azimuthally
extended emission from the BS region becomes clear.
The evolution of the disk brightness distribution after outburst maximum is 
similar to that of the long-period dwarf nova EX~Dra 
(Baptista \& Catal\'an 2001).

\begin{figure}
\includegraphics[bb= 1cm 3cm 19cm 22cm, angle=270, scale=0.38]{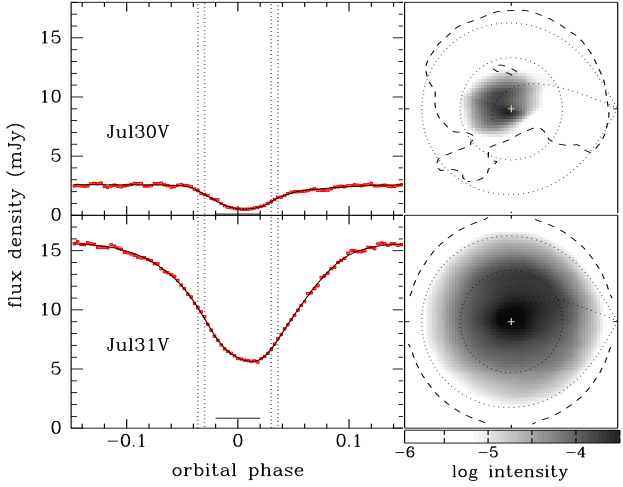}
\caption[c]{V-band light curves and eclipse maps of the 2000 Jul outburst
  at outburst onset (Jul30V) and outburst maximum (Jul31V). 
  The notation is similar to that of Fig.~\ref{fig4}. }
\label{fig5}
\end{figure}

V-band light curves and eclipse maps of the 2000 Jul outburst at outburst
onset (Jul30V) and maximum (Jul31V) are shown in Fig.~\ref{fig5}.  The 
Jul31V light curve and eclipse map are similar to the corresponding B-band 
outburst maximum data. The Jul30V light curve has steeper ingress and 
egress slopes than the Jul30B light curve and the resulting eclipse map
shows an additional light source at disk center aside of the enhanced
emission along the stream trajectory.  
We investigated whether this could be a consequence of underestimating 
the contribution of the WD to the V-band light curve with the following
exercise.  We assumed progressively larger WD contributions to the Jul30V
light curve and computed eclipse maps from the resulting WD-subtracted
light curves (see Sect.~\ref{lightcurve}). The additional contribution 
at disk center is present in the eclipse map even at the limit where the
WD-subtracted light curve starts to show a reverse slope at WD
ingress/egress phases (signaling that too much flux was removed in the
light curve decomposition process).  Thus, we conclude that the extra 
light at disk center in the Jul30V light curve is not related to the WD.
Since there are no V-band observations in the previous nights, it is not 
possible to check whether the reduction in flux of the inner disk regions
observed in the Jul30B B-band map is also present in the V-band.
The tip of the emission pattern along the stream trajectory on Jul30V 
occurs at a radius smaller then the $0.48\;R_{L1}$ quiescent disk radius, 
in agreement with the inferred reduced radius at that date 
(Sect.~\ref{radius}). 
%

\subsubsection{The 2002 Aug outburst}

\begin{figure}
\includegraphics[bb= 1.2cm 3cm 19cm 22cm, scale=0.47]{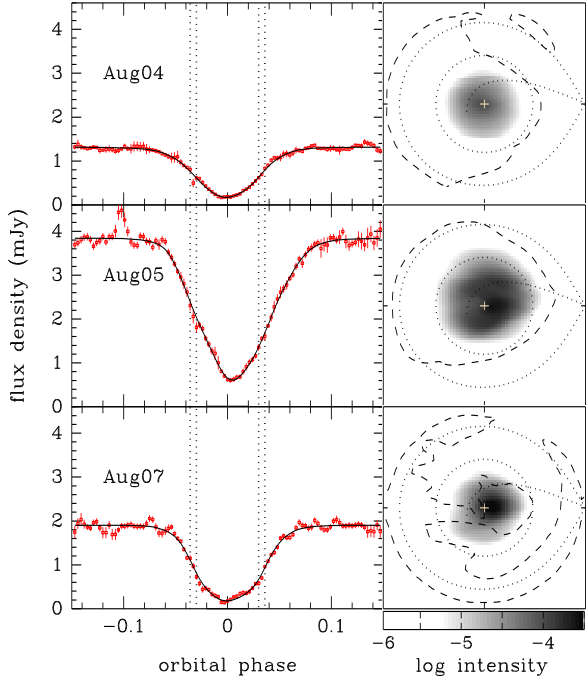}
\caption[c]{B-band light curves and eclipse maps along the 2002 Aug
  outburst.  The notation is similar and the greyscale is the same as 
  that of Fig.~\ref{fig4}.}
\label{fig6}
\end{figure}

B-band light curves and corresponding eclipse maps along the 2002 Aug 
outburst are shown in Fig.~\ref{fig6}. The greyscale of the eclipse maps 
is the same as in Fig.~\ref{fig4}.  This outburst is poorly sampled by the 
amateur astronomers (Fig.~\ref{fig1}) and by our observations 
(Fig.~\ref{fig2}). It is possible to associate the Aug04 data
with the quiescent, pre-outburst stage and the Aug07 data with the late 
decline stage of the short outburst. But it is not clear whether the Aug05 
data corresponds to outburst maximum or not.  Fig.~\ref{fig2} shows that 
there was no perceptible change in out-of-eclipse brightness over the 
$>3$~hr long observing run of that night, in contrast with the clear 
decrease in brightness observed along the observations of the 2000 Jul 
outburst. The constancy of the out-of-eclipse brightness suggests that the 
Aug05 observations framed a fairly stable brightness stage of the outburst 
presumably not too far from outburst maximum.

The symmetric and broad eclipse of Aug04 maps into an axi-symmetric disk 
brightness distribution with no evidence of BS or enhanced emission along 
the gas stream trajectory, in contrast with the 2000 Jul quiescent data.  
Also, the brightness distribution extends up to a smaller radius than in 
2000 Jul.  This eclipse map is reminiscent of that of the 'faint' quiescent 
state of Baptista \& Bortoletto (2004), which may be attributed to a lower 
(long-term) mass transfer rate at that epoch.  
The Aug05 light curve shows an asymmetric V-shaped eclipse with two 
low-amplitude and extended bulges at ingress and at egress.  This leads to
an eclipse map with an elongated annular structure of asymmetric brightness 
distribution superimposed on an axi-symmetric broad baseline distribution.  
This disk brightness distribution bears some resemblance with the eclipse
map of OY~Car on the rise to maximum of a normal outburst (a ring-like
structure, see Rutten et~al. 1992b) and with the He\,II $\lambda\, 4686$
eclipse map of IP~Peg at outburst maximum (a two-armed spiral structure
on top of an extended brightness distribution, see Baptista et~al. 2000).  
The disk radius increases with respect to Aug04 but is far from filling 
the primary Roche lobe.  This seems in accordance with the markedly lower 
amplitude of this outburst. 
The Aug07 light curve has a U-shaped symmetric eclipse. The disk has 
decreased in size back to the pre-outburst radius but the inner disk 
regions remain at the intensity level of the Aug05 map.  The brightness
distribution is skewed towards the L1 point to account for an eclipse
width at half-depth slightly wider than the eclipse of the white dwarf.
%

\subsection{Radial intensity distribution}

\subsubsection{The 2000 Jul outburst}

\begin{figure}
\includegraphics[bb=1cm 1cm 19cm 22cm, scale=0.42]{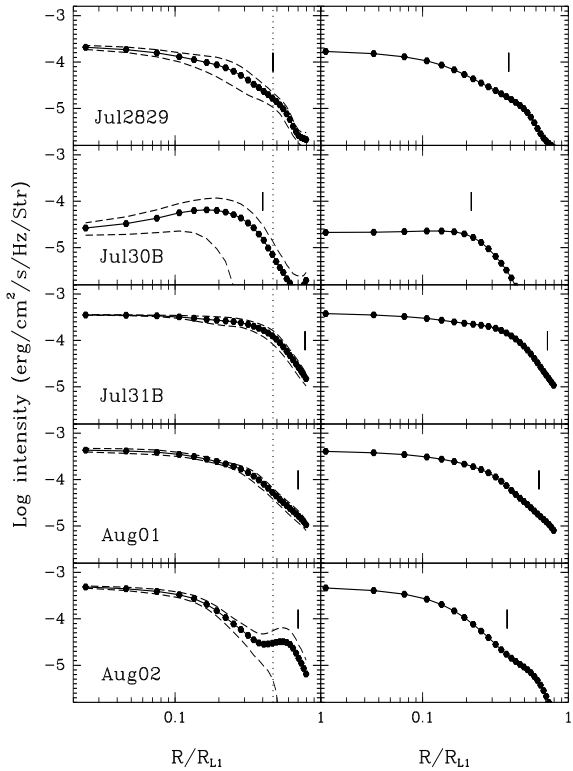}
\caption[c]{The evolution of the B-band radial intensity distribution
  through the 2000 Jul outburst. Left: Average intensity distributions. 
  Dashed lines show the $\pm 1\sigma$ limits on the average intensity. 
  A vertical dotted line indicates the radial position of the BS in 
  quiescence, $R_{bs}=0.48\,R_{L1}$ (Jul2829), while vertical tick marks 
  indicate the radius at which the distribution reaches the BS reference
  intensity level $\log I_{bs}\,= -4.8$.  Right: the symmetric 
  disk-emission distribution, obtained from a cubic spline fit to the
  median of the lower quartile of the radial brightness distribution.  
  Vertical tick marks indicate the radial position at which the disk
  intensity is $\log I_f = -4.77$. }
\label{fig7}
\end{figure}

A more quantitative description of the disk changes during outburst can be 
obtained by analyzing the evolution of the radial intensity distribution. 
The left panels of Fig.~\ref{fig7} illustrate the evolution of the radial 
intensity distribution of the B-band maps along the outburst. We divided 
the eclipse maps in radial bins of $0.03\,R_{L1}$ and computed the median 
intensity at each bin.  These are shown as interconnected circles in 
Fig~\ref{fig7}. The dashed lines show the $\pm 1 \sigma$ limits on the 
average intensity. The labels are the same as in Fig.~\ref{fig4}. The 
large dispersion seen in the intermediate regions of the Jul30B map and
in the outer regions of the Aug02 map reflect the large asymmetries
present in these eclipse maps (gas stream and bright spot, respectively).     
%

In order to quantify the changes in disk size during outburst we defined 
the outer disk radius in each map as the radial position at which the 
intensity distribution is  $\log I_{bs}= -4.8$, which corresponds to the
maximum intensity of the BS in the Jul2829 quiescence eclipse map.  The 
computed outer disk radius is shown as a vertical tick mark in each panel. 
As a reference, the radial position of the quiescent BS is marked by a 
vertical dotted line.    

The disk shrinks from $R_{d}= (0.48 \pm 0.02)\,R_{L1}$ in quiescence to 
$R_{d}=(0.40 \pm 0.11)\,R_{L1}$ at outburst onset (Jul30B), underscoring 
the result of Sect.~\ref{radius}.  The accretion disk then expands and
reaches $R_{d}=(0.78 \pm 0.07)\,R_{L1}$ at outburst maximum (Jul31B).  
The disk decreases in radius to $R_{d}=(0.70 \pm 0.09)\,R_{L1}$ on Aug01
and remains at that radius on the following night (Aug02). 

The DIM predicts that heating and cooling wave-fronts propagate through
the disk during the transitions between the low-viscosity quiescent state
and the high-viscosity outburst state.  The eclipse maps may be used to
measure the movement of transitions fronts. In order to test for the
presence and to trace the movement of transitions fronts in the accretion
disk, we defined an arbitrary reference intensity level $\log I_f= -4.77$. 
This choice is justified by the following arguments. $I_f > I_{bs}$ is 
required in order for it to trace the movement of regions inside the outer 
disk radius.  Furthermore, $I_f$ cannot be much higher than the chosen 
value, otherwise the whole of the brightness distribution on Jul30B would 
be below the reference intensity level.  Choosing other $I_f$ values in the
above range leads to results which are indistinguishable from the ones
present here.

In order to minimize the possible contribution of the BS and gas stream
emission to the disk brightness distribution we calculated the symmetric
disk-emission component. The symmetric component is obtained by slicing 
the disk into a set of radial bins and fitting a smooth spline function
to the median of the lower quartile of the intensities in each bin. 
The spline-fitted intensity in each annular section is taken as the
symmetric disk-emission component. This procedure preserves the baseline
of the radial distribution while removing all azimuthal structure. 
In doing this we are implicitly assuming that the global changes in the 
structure of the accretion disk with time (i.e., transition fronts and 
alike) are roughly axi-symmetric. 
The statistical uncertainties affecting the fitted intensities are 
estimated with the Monte Carlo procedure described in Sect.~\ref{mem}.

The radial position at which the intensity distribution falls below the 
reference intensity level $I_f$ is indicated by vertical tick marks in 
the right panels of Fig.~\ref{fig7}.  It is seen that the radial position 
of the reference intensity changes significantly along the outburst, moving 
outwards on the rising branch and inwards along the decline. These changes 
suggest the propagation of an outward-moving heating wave on the rise and 
of an inward-moving cooling wave on the decline. 

Assuming that the changes in the radial position of the reference intensity
level represent the changes in position of the heating and the cooling
waves, we used the measured positions together with the inferred time
interval between consecutive eclipse maps to estimate the velocities of
the waves along the outburst. The reference time associated to each eclipse
map is the average of the mid-eclipse times of all eclipses included in 
the corresponding median light curve.  The errors in the derived velocities
have contributions from the uncertainties in the time interval between
consecutive eclipse maps and the uncertainties in the measured radial
positions. The inferred speed of the heating wave is $v_{heat} \geq(1.56
\pm 0.05)\,km\,s^{-1}$.  We quote this value as a lower limit because one
cannot discard the possibility that outburst maximum (or largest disk 
radius) occurred earlier than our 2000 Jul 31 observations. The inferred
speed of the inward-moving cooling wave is $v_{cool1} = (-0.30\pm 0.07)
\,km\,s^{-1}$ and $v_{cool2} = (-0.90\pm0.09)\,km\,s^{-1}$, respectively 1
and 2 nights after maximum. The results are summarized in Table~\ref{speed}.
We observe an \emph{acceleration} of the cooling wave as it travels across 
the disk, in contradiction with the predictions of the DIM (Menou et~al. 
1999).  This result is statistically significant at the 6-$\sigma$ 
confidence level.
%

In terms of the $\alpha$ disk formulation of Shakura \& Sunyaev (1973), 
the non-dimensional viscosity parameter of the high state, $\alpha_{hot}$,
can be written as the ratio of the speed at which the heating front 
travels across the disk, $v_{heat}$, and the sound speed inside the
heating front, $c_{s}$ (Lin et~al. 1985; Canizzo 1993),    
\begin{equation}
\alpha_{hot}\approx\frac{v_{heat}}{c_{s}} = 0.082 \left[\frac{v_{heat}}
   {km\,s^{-1}}\right]\left[\frac{T_{f}}{18000\,K}\right]^{-1/2} ,
\end{equation}  
where $T_{f}$ is the temperature of the heating front. Assuming $T_{f}=
18000\,K$ (Menou et~al. 1999), we find $\alpha_{hot}\ga 0.13$, comparable 
to the measured viscosity parameter in quiescence $\alpha_{cool}\simeq 
0.16$ (Baptista \& Bortoletto 2004).  Unrealistically low $T_f$ values 
($< 1000$\,K) are required in order to obtain $\alpha_{hot}$ 
significantly larger than $\alpha_{cool}$.  This result is in contrast 
with the predictions of DIM, which requires that the viscosity in outburst
be significantly larger than in quiescence (i.e., $\alpha_{hot} \ga 5\;
\alpha_{cool}$).

\subsubsection{The 2002 Aug outburst}

\begin{figure}[t]
\includegraphics[bb=0.5cm 5cm 18cm 21cm, scale=0.44]{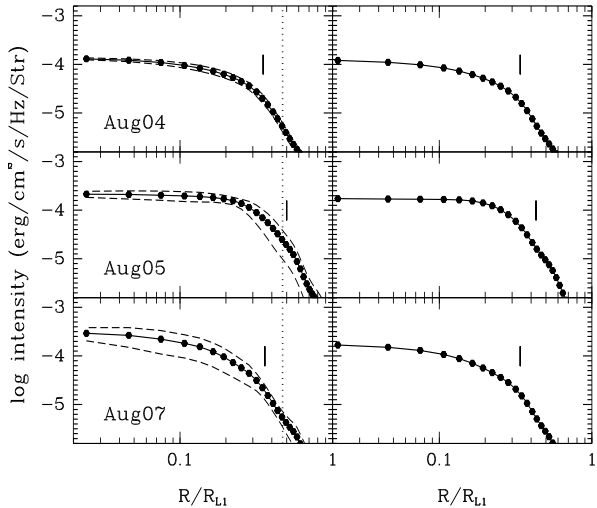}
\caption[c]{Left: radial intensity distributions through the 2002 outburst. 
  The notation is the same as in Fig.~\ref{fig7}. }
\label{fig8}
\end{figure}

The radial intensity distributions along the 2002 Aug outburst are shown
in Fig.~\ref{fig8}. 
%
%
We assumed the same reference intensity levels 
$I_{bs}$ and $I_f$ as before to trace the changes in disk radius and to
estimate the velocity of the heating and cooling waves.  The derived $v_f$
values (and their quoted errors) do not account for the uncertainty in
associating the Aug05 eclipse map with outburst maximum and, therefore,
should be looked at with caution and skepticism. Moreover, there is a
further uncertainty in the measured heating wave speed in this case 
because the Aug04 data does not corresponds to outburst onset but to a
pre-outburst, quiescent state.  
The disk expands from $R_{bs}= (0.35\pm0.02)\,R_{L1}$ in quiescence (Aug04)
to $R_{bs} = (0.50\pm0.08)\,R_{L1}$ at maximum observed brightness (Aug05) 
and shrinks back to the Aug04 radius two nights after that (Aug07).  
We estimated an outward-moving heating wave speed of $v_{heat} \sim 0.31 
\,km\,s^{-1}$ and an inward-moving cooling wave speed of $v_{cool} 
\sim -0.15\,km\,s^{-1}$.

\subsection{Radial temperature distribution} \label{trad}

A simple way to test theoretical disk models is to convert the intensities 
in the eclipse maps to blackbody brightness temperatures, which can then be 
compared to the radial run of the effective temperature predicted by steady 
state, optically thick disk models. It is important to bear in mind that
B- and V-band brightness temperatures are a good approximation to the 
effective temperature only for optically thick disk regions (see, e.g.,
Baptista et~al. 1998b).  Because it is hard to assess the optical depth of 
the disk gas along the outburst with the data at hand, the results from 
this section should be looked upon with some caution.

\begin{figure}
\includegraphics[bb=0.5cm 1cm 17cm 25cm, scale=0.44]{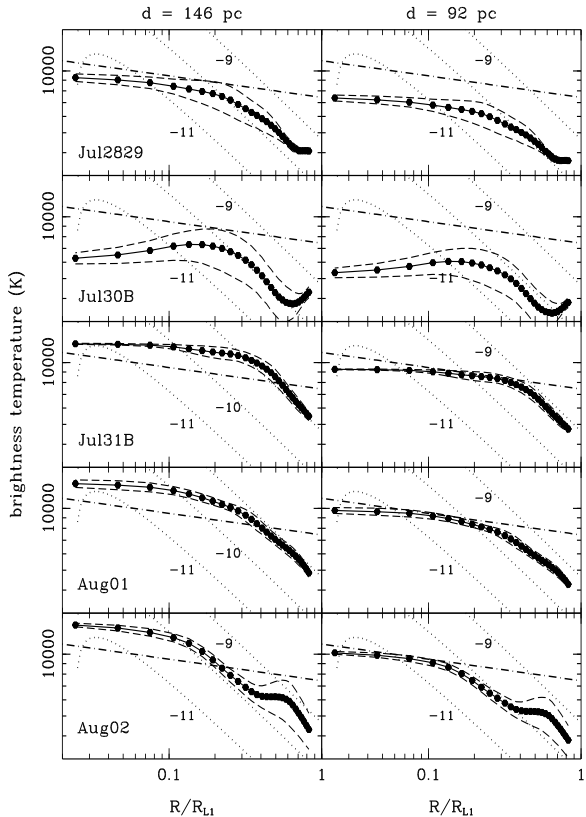}
\caption[c]{The evolution of the radial brightness temperature distribution
  along the 2000 Jul outburst, assuming distances of 146\,pc (left panel)
  and 92\,pc (right) to the binary. Steady state disk models for mass
  accretion rates of \.{M}\,=\,$10^{-9}\,,10^{-10}$ and $10^{-11}\;M_\odot
  \,yr^{-1}$ are plotted as dotted lines for comparison. A dot-dashed line
  marks the critical temperature above which the gas should remain in a
  steady, high mass accretion regime according to DIM. }
\label{fig9}
\end{figure}

\begin{figure}
\includegraphics[bb=3cm 1cm 18.5cm 20cm, angle=270, scale=0.33]{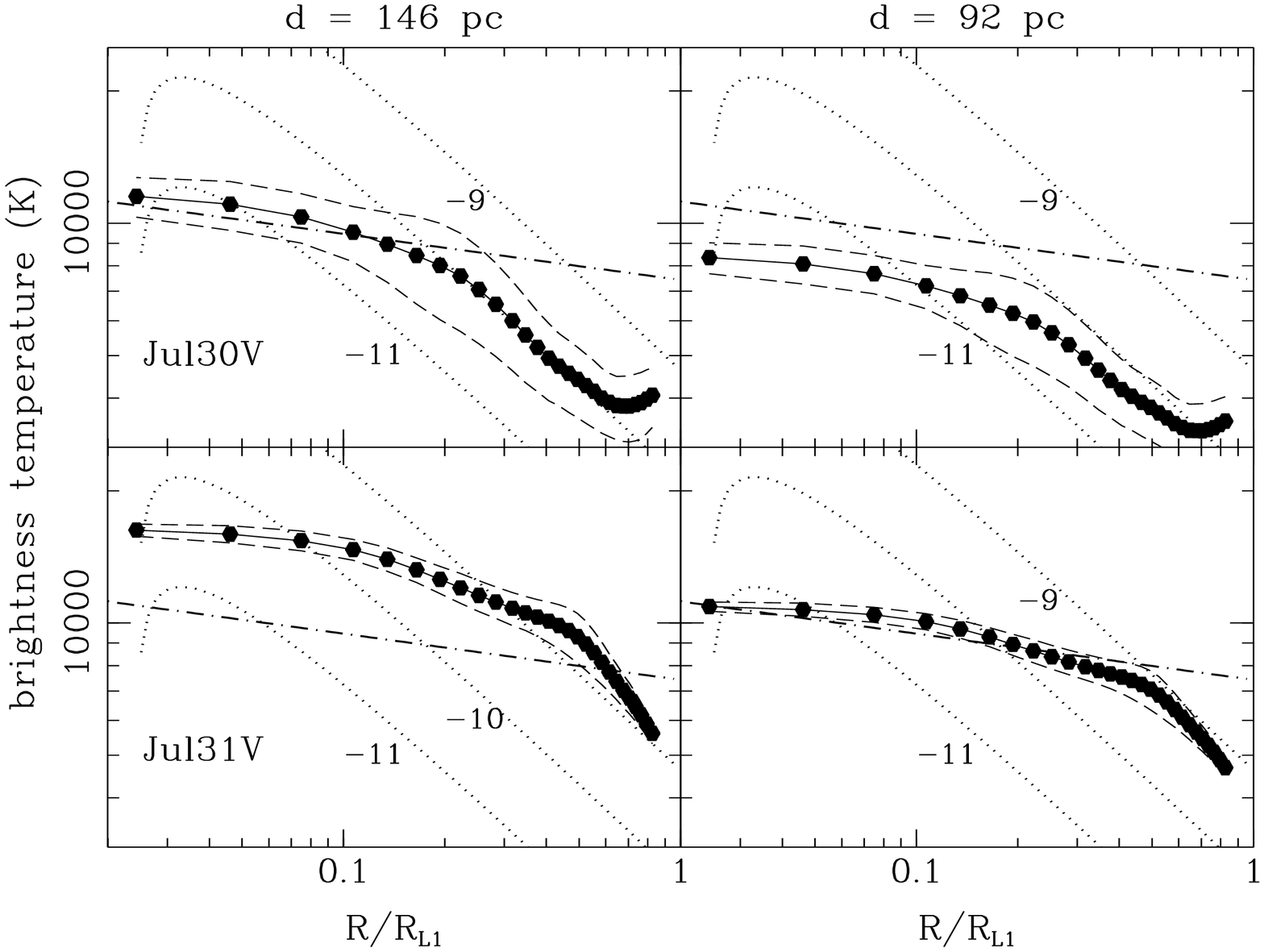}
\caption[c]{The 2000 Jul radial brightness temperature distributions for the 
  V-band data. The notation is the same as in Fig.~\ref{fig9}.}
\label{fig10}
\end{figure}

\begin{figure}
\includegraphics[bb=0.8cm 5cm 18cm 16cm, scale=0.45]{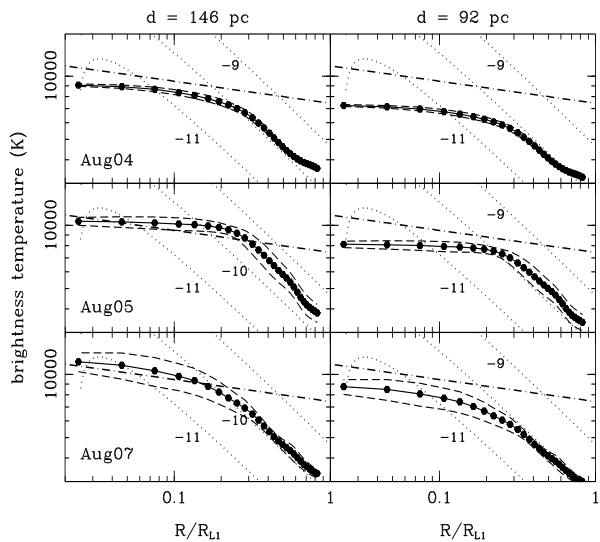}
\caption[c]{Radial brightness temperature distributions for the 2002 Aug
  outburst. The notation is the same as in Fig.~\ref{fig9}. }
\label{fig11}
\end{figure}

Figs.~\ref{fig9}-\ref{fig10} show the evolution of the disk radial 
temperature distribution along the 2000 Jul outburst in a logarithmic 
scale.  The temperature distributions of the 2002 Aug outburst are shown
in Fig.~\ref{fig11}.
%
%
The blackbody brightness temperature that reproduces the observed surface 
brightness at each pixel was calculated assuming distances of 146~pc 
(Vrielmann et~al. 2002a) and of 92~pc (Saito \& Baptista 2006). 
These are respectively shown in the left- and right-hand panels of 
Figs.~\ref{fig9}-\ref{fig11}.
We neglected interstellar reddening, since there is no sign of interstellar
absorption feature at $2200 \AA$ in spectra obtained with the Hubble
Space Telescope (Baptista et~al. 1998a). 
The disk was divided in radial bins of $0.03\,R_{L1}$ and a median
brightness temperature was derived for each bin. These are shown as
interconnected symbols in Figs.~\ref{fig9}-\ref{fig11}. The dashed lines
show the $\pm 1\sigma$ limits on the average temperatures. The larger
$\sigma$-values in the Jul30B, Jul30V and Aug02 distributions reflect the
azimuthal asymmetries caused by the gas stream and BS in these maps. 
Steady-state disk models for mass accretion rates of \.{M}\,=\,$10^{-9}\,,
10^{-10}$ and $10^{-11}\;M_{\odot}\,yr^{-1}$ are plotted as dotted lines
for comparison. These models assume $M_{1} = 0.78\,M_{\odot}$ and a
primary radius of $R_{1}= 0.0103\,R_{\odot}$ (Baptista et~al. 1998a).

The computed brightness temperatures depend on the assumed distance to the
binary.  For a given (fixed) observed flux, the disk is fainter (and 
cooler) if the distance is smaller.  The temperatures and mass accretion
rates quoted in the remainder of this section are for a 92\,pc distance 
to the binary.

Let's first discuss the 2000 Jul outburst distributions.
The quiescent disk (Jul2829) shows a flat temperature distribution for 
$R < 0.3\;R_{L1}$ reminiscent of those seen in other dwarf novae (Wood 
et~al. 1986, 1989). The temperatures range from $\simeq 7000$\,K in the 
inner disk ($R= 0.1\,R_{L1})$ to $\simeq 5000$\,K in the outer disk
regions ($R= 0.4\;R_{L1}$).  
The whole disk brightens at outburst maximum (Jul31B and Jul31V), but the 
outer disk regions become relatively brighter and hotter, leading to a
temperature distribution even flatter than in quiescence for 
$R<0.4\,R_{L1}$. This is in marked contrast with what is seen, e.g., in 
Z~Cha (Horne \& Cook 1985, Wood et~al. 1986) and OY~Car (Rutten et~al. 
1992b), the accretion disks of which transition from a flat brightness 
temperature distribution in quiescence to a steep distribution in good 
agreement with the $T\propto R^{-3/4}$ law in outburst.
Along decline the V2051~Oph inner disk remains at about the same 
temperature ($\sim 10\,000$\,K) while the flat portion of the distribution
progressively recedes towards smaller radii as the outer disk regions cool 
down.
Except for the inner disk regions ($R \la 0.2\;R_{L1}$) -- where the
extra light source result in higher brightness temperatures -- the V-band 
distributions are in agreement with their B-band counterparts, suggesting
that, at least on 2000 Jul 30 and 31, the outer disk was optically thick 
and the derived brightness temperatures are a good approximation to the 
gas effective temperature.

The temperature distributions share a common feature: all show a flat inner 
portion which turns into a steep gradient closely following the $T\propto 
R^{-3/4}$ law in the outer disk regions.  We fitted steady-state disk
models to the steep regions of the temperature distributions to infer
mass accretion rates of \.{M}$= (1.0 \pm 0.1)\times 10^{-10}\;M_\odot\,
yr^{-1}$ in quiescence, $(6.4 \pm 0.1)\times 10^{-10}\;M_\odot\,yr^{-1}$
at outburst maximum, $(3.7 \pm 0.5)\times 10^{-10}\;M_\odot\,yr^{-1}$ on
Aug01, and $(0.9 \pm 0.1)\times 10^{-10}\;M_\odot\,yr^{-1}$ on Aug02.
In the framework of the MTIM, the evolution of the outer disk temperature
distribution could be interpreted as the response of a viscous disk to a
sudden increase in mass transfer rate by a factor of 6.5.  This corresponds
to a brightness increase of $\Delta B= 2$~mag, in line with the observed
2000 Jul outburst amplitude. 

Let's now turn our attention to the 2002 Aug outburst. 
The morphology of the temperature distribution and its evolution with
time are similar to those of the 2000 Jul outburst, although the disk is
cooler at maximum observed brightness (Aug05).  The temperatures in
the quiescent disk (Aug04) are comparable to those of Jul2829.
The temperature distribution is flatter at maximum observed brightness,
with $T\simeq 8\,000$\,K in the inner disk regions (flat portion of the 
distribution) and $T\simeq 5\,700$\,K at $R=0.4\;R_{L1}$. At the late 
decline stage (Aug07) the temperatures in the innermost disk regions ($R 
< 0.15\;R_{L1}$) are still comparable to those at maximum brightness, 
while the outer disk has returned to the temperatures at quiescence.
Fitting steady-state disk models to the steep outer disk regions of the 
temperature distributions one finds mass accretion rates of \.{M}$= (0.8 \pm 
0.1)\times 10^{-10}\;M_\odot\,yr^{-1}$ in quiescence, $(1.7 \pm 0.1)\times 
10^{-10}\;M_\odot\,yr^{-1}$ at maximum observed brightness, and $(0.81 \pm 
0.05)\times 10^{-10}\;M_\odot\,yr^{-1}$ at late decline. This could be
interpreted as the response of a viscous disk to a sudden increase in mass 
transfer rate by a factor of 2.1, corresponding to a brightness increase of
$\Delta B= 0.8$~mag, again in line with the observed 2002 Aug outburst 
amplitude. 
%

Although the nominal temperatures and mass accretion rates change if one 
instead assumes the 146\,pc distance, the qualitative results remain the same.

According to the DIM, there is a critical effective temperature, 
$T_\mathrm{eff}$(crit), below which the disk gas should be while in
quiescence in order to allow the thermal instability to set in, and above
which the disk gas should remain in a steady, high mass accretion regime
(e.g., Warner 1995).  In other words, outbursting accretion disks should be 
hotter than $T_\mathrm{eff}$(crit).  The $T_\mathrm{eff}$(crit) relation is 
plotted as a dot-dashed line in each panel of Figs.~\ref{fig9}-\ref{fig11}.

The distance to the binary is the crucial parameter for the 
$T_\mathrm{eff}$(crit) test.  
If $d=146$\,pc, the evolution of the radial temperature distribution is
consistent with the DIM: the disk temperatures are everywhere below 
$T_\mathrm{eff}$(crit) in quiescence and the outbursting disk shows 
temperatures clearly above that limit.
However, for a 92\,pc distance the outburst occurs at temperatures below 
(2002 Aug) or barely at (2000 Jul) $T_\mathrm{eff}$(crit).
The Aug05 outbursting disk temperatures are systematically lower than 
$T_\mathrm{eff}$(crit) for distances $d\la 120$\,pc.

\subsection{The uneclipsed flux}

The uneclipsed flux is displayed as a horizontal tick centered at phase zero
in the left panels of Figs.~\ref{fig4}, \ref{fig5} and \ref{fig6}.  The 
measured values are listed in Table~\ref{fbg}. We estimated the fractional
contribution of the uneclipsed component to the total flux by computing the
ratio of the uneclipsed flux, $F_\mathrm{un}$, to the average out-of-eclipse
flux level of the corresponding light curve, $F_\mathrm{out}$.  The results
are shown in the third column of Table~\ref{fbg}.
%

The uneclipsed component accounts for a small fraction (1.5 percent) of the 
total system brightness on Jul2829 but increases to 13 percent on Jul30B, the 
highest relative contribution measured.  This is in line with the significant
changes in disk radius and structure observed on Jul30B and corroborates the
idea that this night corresponds to outburst onset.
Although the absolute value of $F_\mathrm{un}$ is larger at outburst maximum, 
its fractional contribution decreases steadily since 2000 Jul 30 and becomes 
negligible on 2000 Aug 2.  The behaviour in the V-band is analogous to that
in the B-band, with a non-negligible $F_\mathrm{un}$ value on Jul30V 
increasing to a maximum (in absolute terms) on the following night. 
The uneclipsed component is negligible throughout the fainter 2002 Aug 
outburst. We draw attention to the fact that the increase in $F_\mathrm{un}$
on the 2000 data precedes outburst maximum by about one day.

The fact that the uneclipsed component is negligible in quiescence and its
variability during outburst indicate that this light does not arise from the
secondary star.  Significant (and variable) uneclipsed components during 
outburst were reported for OY~Car (Rutten et~al. 1992b) and EX~Dra (Baptista
\& Catal\'an 2001).  Two possible explanations to account for the observed 
behaviour would be a variable and largely uneclipsed disk wind or a flaring 
of the accretion disk during outburst.  

Detailed simulations by Wood (1994) show that eclipse maps obtained including
an uneclipsed component or assuming a flared disk may lead to equally good 
fits to the data light curve, making it hard to distinguish between the two
models. Further simulations by Baptista \& Catal\'an (2001) show that modeling
the eclipse of a flared disk with the assumption of a flat disk leads to the
appearance of a spurious uneclipsed component, the flux of which scales with
the disk opening angle, $\theta_d$. 

For a mass accretion rate of $\simeq 10^{-10}\,M_\odot\,yr^{-1}$ 
(sect.~\ref{trad}), we estimate a quiescent disk opening angle of $\theta_d
\simeq 2\degr$ (Meyer \& Meyer-Hofmeister 1982; Smak 1992).
To be accounted for by a flaring of the accretion disk, the observed increase 
in uneclipsed flux by a factor $\simeq 7$ from Jul2829 to Jul30B would demand
an increase in $\theta_d$ by factors of 7-25 (see Baptista \& Catal\'an 2001),
leading to obscuration of a large portion of the accretion disk, in particular
the disk center. This is hard to reconcile with the fact that the white dwarf
at disk center was still visible in the 2000 Jul 30 light curves. Thus,
an outburst-driven disk wind seems the most plausible explanation for the 
observed uneclipsed flux.

\section{Discussion} \label{discuss}

\subsection{The distance to the binary}

Because the radial temperature distribution and the consequent comparison of
disk temperatures with $T_\mathrm{eff}$(crit) depend on the assumed distance
to the binary, we now turn our attention to the distance estimates in the
literature.  

Watts et~al. (1986) found a distance between 90 and 150\,pc assuming a 
steady-state optically thick model for the quiescent accretion disk of 
V2051~Oph.  However, the UV-optical spectrum of V2051~Oph is dominated by 
strong emission lines and a Balmer jump in emission (Watts et~al. 1986, 
Baptista et~al. 1998a) -- indicating important contribution from optically
thin emitting regions -- and the flat temperature distribution of the inner
disk regions in quiescence is in clear disagreement with the 
$T\propto R^{-3/4}$ law of steady-state opaque disks (e.g., Fig.~\ref{fig9}).  
Berriman et~al. (1986) compared the eclipse shape in the H-band with 
computed models to find distances in the range 120-170\,pc. They assumed 
that the red dwarf has a flux density of 1.5\,mJy at the H-band and that 
cool stars have approximately constant surface brightness at the H-band.  
Their models are a poor match to the observed eclipse shape.
Both distance ranges are affected by arguable assumptions and/or too simple
models.

Vrielmann et~al. (2002a) applied a physical parameter eclipse mapping method 
(PPEM) to UBVRI light curves of V2051~Oph in order to derive the spatial 
distribution of temperature and surface density in its accretion disk and
to find a distance of $d=146\pm 20$\,pc to the binary. This method requires
the assumption of an {\em a priori} spectral model for the disk emission 
relating the temperature and surface density to the observed surface 
brightness in the UBVRI passbands. The adopted spectral model is an 
isothermal, pure hydrogen slab of gas in LTE including only bound-free and
free-free H and $\mathrm{H}^-$ emission. 
This model is not adequate to handle emission from disk atmospheres with 
vertical temperature gradients. In particular, it might fail to model a 
temperature inversion in the disk photosphere (i.e., a hot disk chromosphere
with strong emission lines), as it seems to be the case in V2051~Oph 
(Vrielmann et~al. 2002a; Saito \& Baptista 2006) and other quiescent dwarf
novae. The application of this method and model to the dwarf nova HT~Cas 
leads to a distance 50 per cent larger then that inferred from the Na\,I 
and TiO absorption spectrum of the secondary star (Vrielmann et~al. 2002b). 
The PPEM distance to V2051~Oph may be overestimated as well. 
Moreover, Vrielmann et~al. (2002a) mistakenly adopted a WD radius of $R_1= 
0.0244\;R_\odot$, a factor of 2.4 larger than the $R_1= 0.0103\;R_\odot$
value quoted by Baptista et~al. (1998a). As a result, the WD ingress and 
egress features in the light curve become artificially longer and the PPEM
code associates more surface flux to the WD than it should.  The WD 
accounts for a significant fraction of the V2051~Oph optical light in 
quiescence (see Fig.\,3 of Vrielmann et~al. 2002a) and dominates the 
procedure used to estimate the distance.  The extra WD surface flux has to
be compensated for by increasing the distance to the binary.

We are therefore left with the $d=92^{+30}_{-35}$\,pc distance estimate of
Saito \& Baptista (2006). This was obtained by fitting white dwarf atmosphere 
models of solar composition to the extracted UV-optical WD spectra at the
time the WD was most clearly seen in V2051~Oph (Baptista et~al. 1998a; Saito
\& Baptista 2006). In the absence of a more robust determination we take
this as the best available distance estimate to V2051~Oph.

\subsection{The cause of the V2051~Oph outbursts}

Eclipse mapping analysis along two outburst cycles in V2051~Oph allows us
to test the predictions of dwarf novae outburst models listed in 
Sect.~\ref{intro}. Here we discuss the cause of the outbursts of V2051~Oph
under the light of the results presented in the previous sections.

MTIM predicts that the disk shrinks at outburst onset (e.g., Warner 1995
and references therein). 
We found that the disk decreased in brightness and radius at the night 
before the 2000 Jul outburst maximum while the uneclipsed light in the 
B-band increased (from a negligible contribution) to 13 per cent of the
total light.  It is hard to avoid the interpretation that the changes in
disk structure and uneclipsed flux are a consequence of the outburst onset
and the following conclusion that our result confirms the MTIM prediction.

DIM requires that $\alpha_{hot} \ga 5\; \alpha_{cool}$ (e.g., Lasota 2001). 
In MTIM, there is no reason for the viscosity parameter to change from
quiescence to outburst. We measure $\alpha_{hot} \simeq \alpha_{cool}
\simeq 0.13$, confirming another MTIM prediction.
In favor of DIM, one could argue that the true $v_{heat}$ may be much larger
than measured, contributing to increase $\alpha_{hot}$ relative to 
$\alpha_{cool}$.  However, there is no observational support for an 
$\alpha_{hot}$ significantly larger than inferred. Vrielmann \& Offutt (2003)
measured a heating wave speed of $v_{heat} = +1.8\,km\,s^{-1}$ during a
superoutburst in V2051~Oph, in good agreement with our result.
One could alternatively dismiss the $\alpha_{cool}$ 
measurement of Baptista \& Bortoletto (2004), leaving room for a much smaller
$\alpha_{cool}$ value.  Nevertheless, the good agreement between the radial
temperature distribution of the outer disk and the $T\propto R^{-3/4}$ law
suggests that the outer regions of the V2051~Oph accretion disk are in a
steady-state both in outburst and in quiescence, implying that the disk
viscosity there is always high.

DIM predicts that the cooling wave decelerates as it travels towards disk
center (Menou et~al. 1999). In MTIM, the measured $v_{cool}$ reflects 
the local viscous timescale and may increase or decrease depending on the
radial dependence of the viscosity parameter $\alpha$. We find that the
cooling wave accelerates as it travels towards disk center, in clear
disagreement with the DIM prediction.

DIM predicts that an outbursting accretion disk is hotter than 
$T_\mathrm{eff}$(crit) (e.g., Warner 1995 and references therein). 
There is no quiescence/outburst temperature restriction in MTIM.
For an assumed distance of 92\,pc, we find that the whole of the 
low-amplitude 2002 Aug outburst occurs at temperatures below 
$T_\mathrm{eff}$(crit), and that the hottest parts of the accretion disk 
during the normal 2000 Jul outburst barely reach $T_\mathrm{eff}$(crit). 
The 2002 Aug outburst fails to meet the $T > T_\mathrm{eff}$(crit) 
requirement for distances $d<120$\,pc.  The morphology and time evolution
of the radial temperature is similar on both outbursts, and there is no
reason to believe they are powered by different physical mechanisms. If 
one admits the 2002 Aug outburst is not driven by a thermal-viscous disk
instability, the same conclusion must be applied to the normal 2000 Jul
outburst. The inconsistency between DIM and the observations here could
be alleviated if one argues that the V2051~Oph accretion disk is not optically
thick during outbursts and that the derived brightness temperatures
underestimate the disk gas effective temperatures.

If taken separately, none of the individual results would be enough
to discard DIM as a powering mechanism for the outbursts of V2051~Oph.
Nevertheless, when taken in combination the above results make a strong
case against DIM and in favour of MTIM as the cause of the outbursts in 
V2051~Oph.  All evidences discussed here lead to the conclusion that the
outbursts of V2051~Oph are driven by episodes of enhanced mass-transfer
from its secondary star.

Before one claims that V2051~Oph cannot be considered a 'true' dwarf nova, 
we remind it is an SU~UMa star, showing not only 2~mag amplitude normal
outbursts, but also longer and brighter superoutbursts (Kiyota \& Kato
1998; Vrielmann \& Offutt 2003).
It is also worth noting that finding a dwarf nova the outbursts of which
are not powered by disk instabilities does not necessarily kills DIM, but
warns that the current view about dwarf nova outbursts must be revised to
include the possibility that both outburst mechanisms coexist, perhaps
on different dwarf nova subtypes that have not been clearly distinguished
up to this point.

We propose that the viscosity in the accretion disk of V2051~Oph is always
high, independent of the disk mass inflow rate. 
In this scenario, there is no room for matter to accumulate in the outer
disk regions nor for thermal-viscous instabilities to set in. 
Because of the low-mass transfer rate and the efficient accretion engine, 
the disk has low densities.
A low density quiescent disk enables the (denser) infalling gas to 
'overflow' the outer disk regions, creating a bright gas stream along the 
ballistic trajectory ahead of the disk rim (cf., Baptista \& Bortoletto 2004) 
and making the trailing lune of the disk significantly brighter than the 
leading lune -- as observed by Watts et~al. (1986).
This is supported by numerical simulations of accretion discs, which show
that when the infalling gas stream is significantly denser than the initial 
disc gas, the stream penetrates the disc -- allowing matter to be deposited
at the inner disc regions -- and no bright spot forms at disk rim 
(Bisikalo et~al. 1998a,b; Makita et~al. 2000).
Because the energy released at the disk rim is small in comparison to that
released along the stream closer to the WD, there is no pronounced orbital 
hump caused by anisotropic BS emission in the light curve -- in contrast with
the conspicuous orbital hump seen in the light curves of the low-viscosity
disk dwarf novae OY~Car (e.g., Wood et~al. 1989), Z~Cha (e.g., Wood et~al. 
1986) or IP~Peg (e.g., Wood \& Crawford 1986). 
Furthermore, without strong anisotropic BS emission, there will be no 
observable increase in BS luminosity to signal a sudden enhancement of mass
transfer rate at outburst onset. Since the transferred matter is allowed to
flow all the way down to the inner disk regions (i.e., at the circularization
radius), a burst of enhanced mass-transfer from the secondary star can lead
to inside-out outbursts similar to that of 2000 Jul.
Variability in the mass transfer from the secondary star is therefore 
responsible not only for the strong flickering and the long-term brightness
changes (Baptista \& Bortoletto 2004) but also for the infrequent normal
(increase in \.{M}$_2$ by factors 5-10) and low-amplitude (increase in 
\.{M}$_2$ by factors 2-3) outbursts.
Going one step further, there is no reason to believe the disk would
switch to a low-viscosity state just before a superoutburst.  Therefore,
this scenario implies that the rare superoutbursts of V2051~Oph are also
powered by (stronger) bursts of enhanced mass transfer from its secondary
star.

The increase in the luminosity of the BS at outburst onset and the outburst 
type issue (inside-out versus outside-in) are no valid tests to distinguish
between DIM and MTIM in dwarf novae such as V2051~Oph.  
Are there other, useful tests to distinguish between the two outburst models? 
Section~1 gives a list of observational tests for which the predictions of 
both models are quite different (and, therefore, relatively easy to
distinguish). Unfortunately, they are only applicable to the restricted 
group of eclipsing dwarf novae. 
Comparing the disc radius in quiescence and at outburst onset would be the 
most straightforward test, but also the hardest to perform because it 
requires observations framing few hours around the highly unpredictable 
start of the outburst.
The comparison of the disc brightness temperature distribution with the
$T_\mathrm{eff}$(crit) relation requires good knowledge of the distance to
the binary and depends on the assumption that the brightness temperature 
is a good approximation to the disc gas effective temperature (or, 
alternatively, on the availability of an adequate model relating both
temperatures).
Tracing changes in speed of the cooling wave along decline is probably the
less demanding test from the observational point of view -- one needs good 
sampling of the decline branch of the outburst -- but may be inconclusive
if one finds that the cooling wave decelerates.
Because $\alpha_{hot}$ is easily inferred from the timescale of the outburst, 
the key aspect of the comparison of $\alpha_{hot}$ and $\alpha_{cool}$ is 
the measurement of the viscosity parameter in quiescence. 
A promising step in this regard may be to perform spatially-resolved 
flickering studies of quiescent dwarf novae. The identification of a disc 
flickering component allows an estimate of the magnitude and the radial 
dependency of the disc viscosity (e.g., Baptista \& Bortoletto 2004).

The present results raise a set of exciting and challenging questions:
How many other high-viscosity disk dwarf novae similar to V2051~Oph are there?
What fraction of the dwarf novae class do they account for?
Why do their accretion disks have permanent high viscosity?
Why and how the secondary stars of CVs change their mass transfer rates
by large amounts on short, 1-2 days timescales?

\acknowledgments

R.B. is grateful to the chilean CV community for interesting discussions
about the results of this work.
We thank an anonymous referee for useful comments which helped to improve
the presentation of our results.
This paper was based on observations made at Laborat\'{o}rio Nacional de
Astrof\'{i}sica/CNPq, Brazil. In this research we have used, and acknowledge
with thanks, data from the AAVSO International Database that are based on
observation collected by variable star observers worldwide. 
This work was partially supported by CNPq/Brazil through research grant
62.0053/01-1-PADCT III/Milenio. R.B. acknowledges financial support from
CNPq/Brazil through grants 300.354/96-7, 301.442/2004-5 and 200.942/2005-0.
R.F.S. acknowledges financial support from a CNPq/Brazil undergraduate
research fellowship.

\clearpage

%
\begin{deluxetable}{ccrcccrcc}
\tablecaption{Journal of the observations\label{observa}}
\tablewidth{0pt}
\tablehead{
\colhead{date} & \colhead{HJD start} & \colhead{$\Delta\,t$} & \colhead{band}
& \colhead{E \tablenotemark{a}} & \colhead{phase} & \colhead{$N_p$} & 
\colhead{Quality \tablenotemark{b}} & \colhead{brightness} \\
& \colhead{(2.450.000+)} & \colhead{(s)} && \colhead{(cycle)} & 
\colhead{range} &&& \colhead{state}
}
\startdata
2000 Jul 28 & 1754.4464 &  5 &  B & 136293 & $-0.19,+0.50$ &  726 &  A & 
quiescence \\
 & 1754.4894 &  5 &  B & 136294 & $-0.50,+0.50$ &  941 &  A & quiescence \\
 & 1754.5518 &  5 &  B & 136295 & $-0.50,+0.50$ & 1074 &  A & quiescence \\ 
 & 1754.6142 &  5 &  B &(136296)& $-0.50,+0.23$ &  538 &  A & quiescence \\ 
[0.5ex]

2000 Jul 29 & 1755.3952 &  5 &  B & 136308 & $+0.01,+0.50$ &  281 &  A & 
quiescence \\
 & 1755.4257 &  5 &  B & 136309 & $-0.50,+0.50$ & 1079 &  A & quiescence \\
 & 1755.4882 &  5 &  B & 136310 & $-0.50,+0.50$ & 1078 &  A & quiescence \\
 & 1755.5506 &  5 &  B & 136311 & $-0.50,+0.30$ &  860 &  A & quiescence \\
 & 1755.6176 &  5 &  B & 136312 & $-0.43,+0.36$ &  848 &  A & quiescence \\ 
[0.5ex]

2000 Jul 30 & 1756.4169 &  5 &  B & 136325 & $-0.62,+0.29$ &  986 &  B & 
quiescence \\
 & 1756.5029 &  5 &  V & 136326 & $-0.24,+0.50$ &  793 &  B & quiescence \\
 & 1756.5494 &  5 &  V & 136327 & $-0.50,+0.30$ &  836 &  B & quiescence \\
 & 1756.6285 &  5 &  V & 136328 & $-0.23,+0.45$ &  850 &  C & quiescence \\ 
[0.5ex]

2000 Jul 31 & 1757.4083 &  5 &  V & 136341 & $-0.74,+0.50$ & 1296 &  C & 
outburst \\
 & 1757.4859 &  5 &  V &(136342)& $-0.50,-0.09$ &  376 &  B & outburst \\
 & 1757.5116 &  5 &  B & 136342 & $-0.09,+0.50$ &  627 &  A & outburst \\
 & 1757.5483 &  5 &  B & 136343 & $-0.50,+0.57$ & 1135 &  A & outburst \\
 & 1757.6152 &  5 &  V & 136344 & $-0.43,+0.45$ &  949 &  A & outburst \\ 
[0.5ex]

2000 Aug 1 & 1758.4741 &  5 &  B & 136358 & $-0.67,+0.50$ &  420 &  B & 
decline \\
 & 1758.5472 & 10 &  B & 136359 & $-0.49,+0.50$ &  254 & C & decline \\
 & 1758.6096 & 15 &  B & 136360 & $-0.50,+0.14$ &  139 & C & decline \\ 
[0.5ex]

2000 Aug 2 & 1759.4485 &  5 &  B & 136373 & $-0.06,+0.50$ &  202 &  A & 
decline \\
 & 1759.4837 &  5 &  B & 136374 & $-0.50,+0.50$ &  323 &  B & decline \\
 & 1759.5461 &  5 &  B & 136375 & $-0.50,+0.10$ &  217 &  A & decline \\ 
[1ex]

2002 Aug 4 & 2491.4401 & 10 &  B & 148099 & $-0.66,+0.50$ & 626 & B &
quiescence \\
 & 2491.5126 & 10 &  B & 148100 & $-0.50,+0.75$ &  672& B & quiescence \\ 
[0.5ex]

2002 Aug 5 &  2492.4126  &10 &  B & 148114 & $-0.08,+0.50$ &  315 & A & 
outburst \\ 
 & 2492.4491  &10 &  B &(148115)& $-0.50,+0.25$ &  255 & C & outburst \\
 & 2492.5278  &10 &  B & 148116 & $-0.24,+0.16$ &  174 & C & outburst \\
[0.5ex]

2002 Aug 7 & 2494.4529  &20 &  B & 148147 & $-0.40,+0.58$ &  266 & A & 
decline \\ [0.5ex]
\enddata
\tablenotetext{a}{ with respect to the ephemeris of eq.~(\ref{efem}).}
\tablenotetext{b}{ sky conditions: A= photometric (main comparison stable), 
 B= good (some sky variations), C= poor (large variations and/or clouds).} 
\end{deluxetable}

\clearpage 

%
\begin{deluxetable}{lccccc}
\tablecaption{Measuring the speed of transitions waves\label{speed}}
\tablewidth{0pt}
\tablehead{
\colhead{eclipse map} & \colhead{$R_f/R_{L1}$} & \colhead{$\Delta R_f/R_{L1}$}
& \colhead{$\Delta t$} & \colhead{$v_f$} \\
\colhead{sequence} &&& \colhead{(days)} & \colhead{($km\,s^{-1}$)}
}
\startdata
Jul30B\,$\mapsto$\,Jul 31B 
& $0.22\,\mapsto\,0.72$  & $+0.50$ & 1.09 & $+1.56 \pm 0.05$ \\
Jul31B\,$\mapsto$\,Aug 01  
& $0.72\,\mapsto\,0.63$  & $-0.09$ & 1.03 & $-0.30 \pm 0.07$ \\
Aug01\, $\mapsto$\,Aug 02  
& $0.63\,\mapsto\,0.38$  & $-0.25$ & 0.94 & $-0.90 \pm 0.09$ \\ [1ex]
Aug04\, $\mapsto$\,Aug 05  
& $0.34\,\mapsto\,0.43$  & $+0.09$ & 0.97 & $+0.31 \pm 0.04$ \\ 
Aug05\, $\mapsto$\,Aug 07  
& $0.43\,\mapsto\,0.34$  & $-0.09$ & 2.00 & $-0.15 \pm 0.02$ \\ 
\enddata
\tablecomments{ $R_f$ is the radius at which the intensity falls below 
  $\log I_f = -4.77$; $\Delta t$ is the time interval between the two 
  consecutive eclipse maps; $v_{f}=(\Delta R_{f}/\Delta t)$ is the speed 
  of the moving wave (in $km\,s^{-1}$).}
\end{deluxetable}

%
\begin{deluxetable}{lcc}
\tablecaption{The uneclipsed flux\label{fbg}}
\tablewidth{0pt}
\tablehead{
\colhead{Map} & \colhead{$F_\mathrm{un}$\,(mJy)} & 
\colhead{$F_\mathrm{un}/F_\mathrm{out}$}
}
\startdata
 Jul2829 & $0.03 \pm 0.02$ & 0.015 \\ 
 Jul30B  & $0.20 \pm 0.04$ & 0.13 \\ 
 Jul31B  & $0.42 \pm 0.07$ & 0.04 \\ 
 Aug01   & $0.17 \pm 0.06$ & 0.02 \\
 Aug02   & $0.00 \pm 0.01$ &  -- \\ [1ex]

 Jul30V  & $0.10 \pm 0.02$ & 0.04 \\
 Jul31V  & $0.62 \pm 0.06$ & 0.04 \\ [1ex]

 Aug04   & $0.01 \pm 0.01$ & -- \\
 Aug05   & $0.02 \pm 0.01$ & -- \\
 Aug07   & $0.00 \pm 0.01$ & -- \\ 
\enddata
\end{deluxetable}

\end{document}